\RequirePackage{lineno}
\documentclass[10pt,numberedappendix]{emulateapj}
\pdfoutput=1
\usepackage{comment}
\usepackage{graphicx}
\usepackage{color}
\usepackage{amsmath}
\usepackage{amsfonts}
\usepackage{soul}

\begin{document}

\title{
Observational Selection Effects with Ground-based Gravitational Wave Detectors
}

\author{Hsin-Yu Chen\altaffilmark{1}, Reed Essick\altaffilmark{2}, Salvatore
  Vitale\altaffilmark{2}, Daniel E. Holz\altaffilmark{1}, Erik Katsavounidis\altaffilmark{2}}

\altaffiltext{1}{University of Chicago}
\altaffiltext{2}{MIT LIGO Laboratory}

\begin{abstract}
Ground-based interferometers are not perfectly all-sky instruments, and it is important to account for their behavior when considering the distribution of detected events.
In particular, the LIGO detectors are most sensitive to sources above North America and the Indian Ocean and, as the Earth rotates, the sensitive regions are swept across the sky.
However, because the detectors do not acquire data uniformly over time, there is a net bias on detectable sources' right ascensions.
Both LIGO detectors preferentially collect data during their local night; it is more than twice as likely to be local midnight than noon when both detectors are operating.
We discuss these selection effects and how they impact LIGO's observations and electromagnetic follow-up.
Beyond galactic foregrounds associated with seasonal variations, we find that equatorial observatories can access over $80\%$ of the localization probability, while mid-latitudes will access closer to $70\%$.
Facilities located near the two LIGO sites can observe sources closer to their zenith than their analogs in the South, but the average observation will still be no closer than $44^\circ$ from zenith.
We also find that observatories in Africa or the South Atlantic will wait systematically longer before they can begin observing compared to the rest of the world, although there is a preference for longitudes near the LIGOs.
These effects, along with knowledge of the LIGO antenna pattern, can inform electromagnetic follow-up activities and optimization, including the possibility of directing observations even before gravitational-wave events occur.
\end{abstract}

\maketitle


\section*{introduction}\label{s:introduction}

The detection of binary black holes with the Laser Interferometer Gravitational-wave Observatory~\citep[LIGO;][]{0264-9381-32-7-074001,detectionPRL,2016PhRvL.116x1103A} has ushered in the age of gravitational wave astronomy. 
A particularly promising avenue for exploring new physics is multi-messenger astronomy, involving the joint detection of GW sources, electromagnetic (EM) signals, or astrophysical particles~\citep{Kulkarni,Gehrels}.
During the last years of initial LIGO and Virgo, as well as in advanced LIGO's first observing run~\citep[O1;][]{O1BBH}, a large consortium of electromagnetic (EM) observers followed up GW candidates~\citep{emfollow,emS6,emS6lowlat}, and there has been substantial effort to plan and optimize EM follow-up. 
Previous work compared the advantages and disadvantages of different telescopes~\citep{KandN}.
Regardless of a facility's hardware, observatories at different locations will have systematically different opportunities to follow-up GW events due to properties of the GW detector network.

Even if astrophysical sources are distributed isotropically on the sky, GW detections with the two LIGO detectors will not be.
Because of the detectors' locations (Hanford, WA and Livingston, LA) and duty cycles, detectable GW sources preferentially come from certain locations on the celestial sphere, and the preferred regions vary with the seasons. 
Follow-up of EM counterparts with emission timescales of less than a few weeks will be especially affected by these biases.
In particular, the Earth's rotation limits ground-based EM follow-up facilities.
Different sites can access different parts of the GW localization maps at different times and therefore have different expectations for the fraction of counterparts they can detect.
These GW selection effects have other important ramifications for follow-up efforts, such as the amount of time before an average source will be accessible and the average air mass expected.

GW detectors do not operate continuously~\citep{S6Detchar,O1Detchar}.
Previous studies considered their duty cycle when estimating detection rates~\citep{F2Y, ObservingScenarios, BBHrate}.
However, their operation is not uniformly distributed in time, and instead shows a strong preference for acquiring data during their local night.
We describe this behavior quantitatively and assess its influence on GW detections as well as EM follow-up.

We first explain the sources of biases and their impact on GW detections in \S~\ref{s:gw}. 
We show the effects of the bias on ground-based EM follow-up facilities in
\S~\ref{s:em}, and conclude in \S~\ref{s:discussion}.


\section{observational bias introduced by ground-based gravitational wave detectors}\label{s:gw}

Ground-based GW interferometers do not have isotropic sensitivity. 
Over time, the antenna pattern produces systematic preferences for the locations of detectable sources.
These can be split into a dependence on the source's declination, which will not improve until additional detectors are added to the network, and a dependence on the source's right ascension, which can be mitigated with existing facilities. 

\begin{figure*}
    \includegraphics[width=1.0\textwidth]{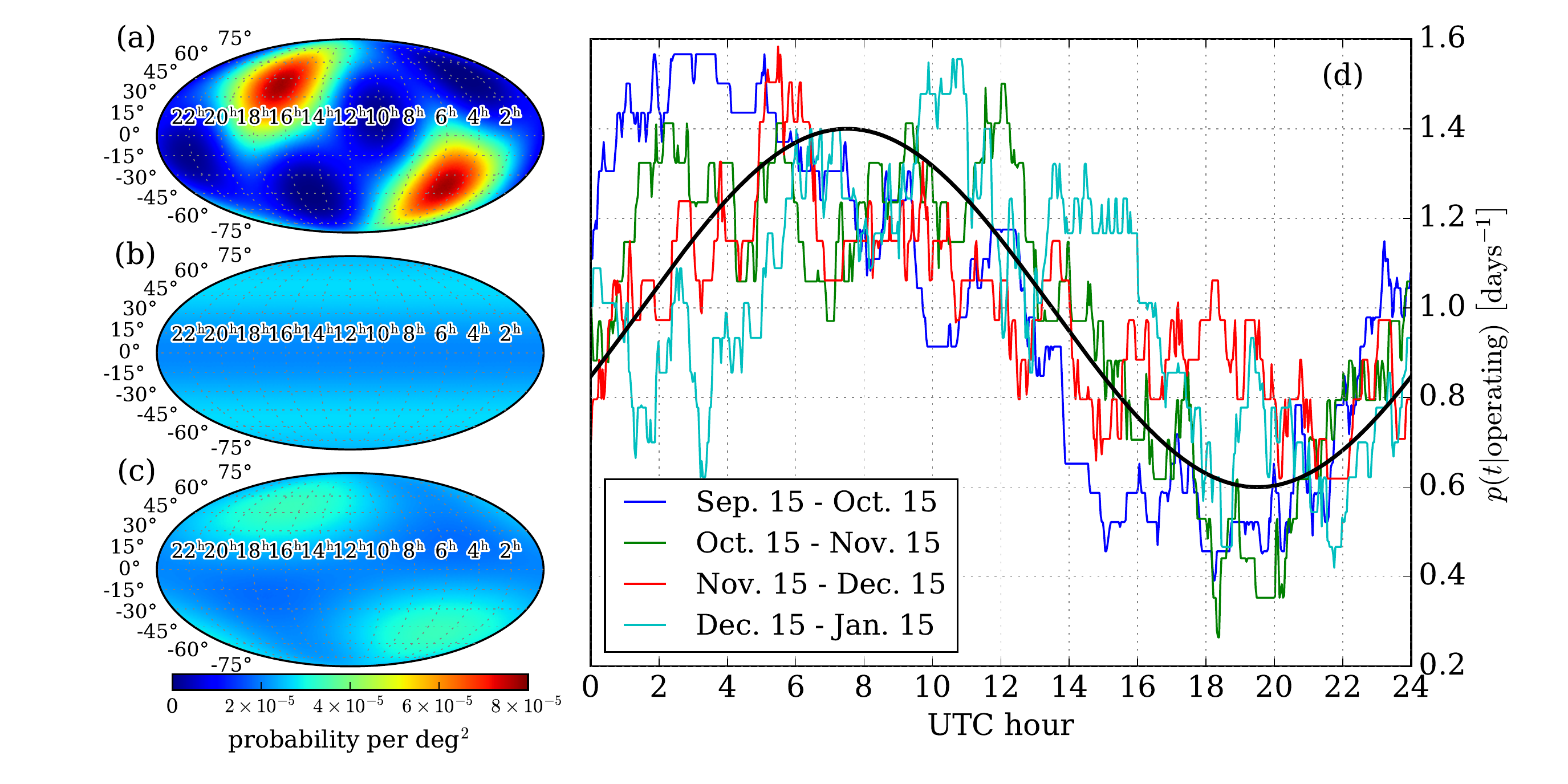}
    \caption{
        (a) The LIGO Hanford and Livingston network antenna pattern in equatorial coordinates at 00:29:18 UTC on Sep. 14, 2015. The maxima lie above North America and the Southern Indian Ocean.
        (b) The antenna pattern swept over the celestial sphere assuming uniform operation throughout time, resulting in two maxima bands in mid-declinations.
        (c) The antenna pattern swept over the celestial sphere assuming a typical diurnal cycle. 
        This produces a dependence on right ascension that persists over timescales of a few days to weeks.
        (d) The observed operation of the LIGO detectors during O1 and a sinusoidal model as a function of UTC time. 
        We require both LIGO detectors to be operating at the same time, although each detector shows similar behavior individually.
        }
    \label{f:gwbias}
\end{figure*}


\subsection{Dependence on Declination}\label{s:declination}

Because of projection effects, GW detectors are most sensitive to signals coming from above or below the plane defined by their arms~\citep{thorne300}.
The two LIGO detectors, for example, are most sensitive, and equally sensitive, to sources directly above North America and above the Indian Ocean (Fig \ref{f:gwbias}a).
As the Earth rotates, the antenna pattern is swept across the celestial sphere, creating preferred bands in the mid-latitudes of both hemispheres (Fig \ref{f:gwbias}b). 
This is determined primarily by the detectors' relative sensitivity and the geometry of the network. 
For the two LIGO detectors, we expect more detections at mid-declinations compared to low- or high-declinations.


\subsection{Dependence on Right Ascension}\label{s:right ascension}

The antenna pattern also introduces a dependence on longitude, which translates to a dependence on the right ascension. 
If the detectors operated uniformly in time, then this right ascension dependence would average away as the Earth rotates. 
However, because the duty cycle is not uniform in time, this engenders a net bias in favor of certain right ascensions. 

Ground-based GW detectors' data acquisition shows a clear diurnal cycle.
In previous runs (S6~\citep{PhysRevD.85.082002, PhysRevD.87.022002} and earlier) this was due primarily to anthropogenic noise; the detectors observed lower ambient noise at night because humans were less active.
During O1, we observed a similar diurnal cycle. 
Although anthropogenic noise was mitigated by improved seismic isolation, commissioning activities still preferentially occurred during the day at the detector sites.
This, combined with a modest duty cycle~\citep{2016PhRvL.116m1103A}, generated a non-trivial preference for acquiring data during the sites' night (Fig.~\ref{f:gwbias}d).
We model this preference with a probability distribution for the time of day during which both interferometers are likely to be operating:
\begin{equation}\label{e:bias}
    p(t|\text{operating}) = \frac{1}{\text{day}}\left( 1 + A \sin\left(\frac{2 \pi t}{\text{day}} - \frac{\pi}{8}\right) \right)
\end{equation}
The amplitude ($A$) reflects the extent of the day/night bias and is typically near 0.4.
This means the detectors are 2.3 times more likely to record data at their local midnight than at noon. 
We note that this bias may be decreased by reducing the amount of commissioning activity during the day or by increasing the overall duty cycle of the instruments (e.g., reducing the duration of down-time due to events uncorrelated with the diurnal cycle). 
As the duty cycle increases, the relative importance of the diurnal behavior will decrease.\footnote{Please see Sec S3 of the Appendix for more details.}
However, we always expect some small diurnal cycle to be present.

\begin{figure*}[b!]
    \includegraphics[width=1.0\textwidth]{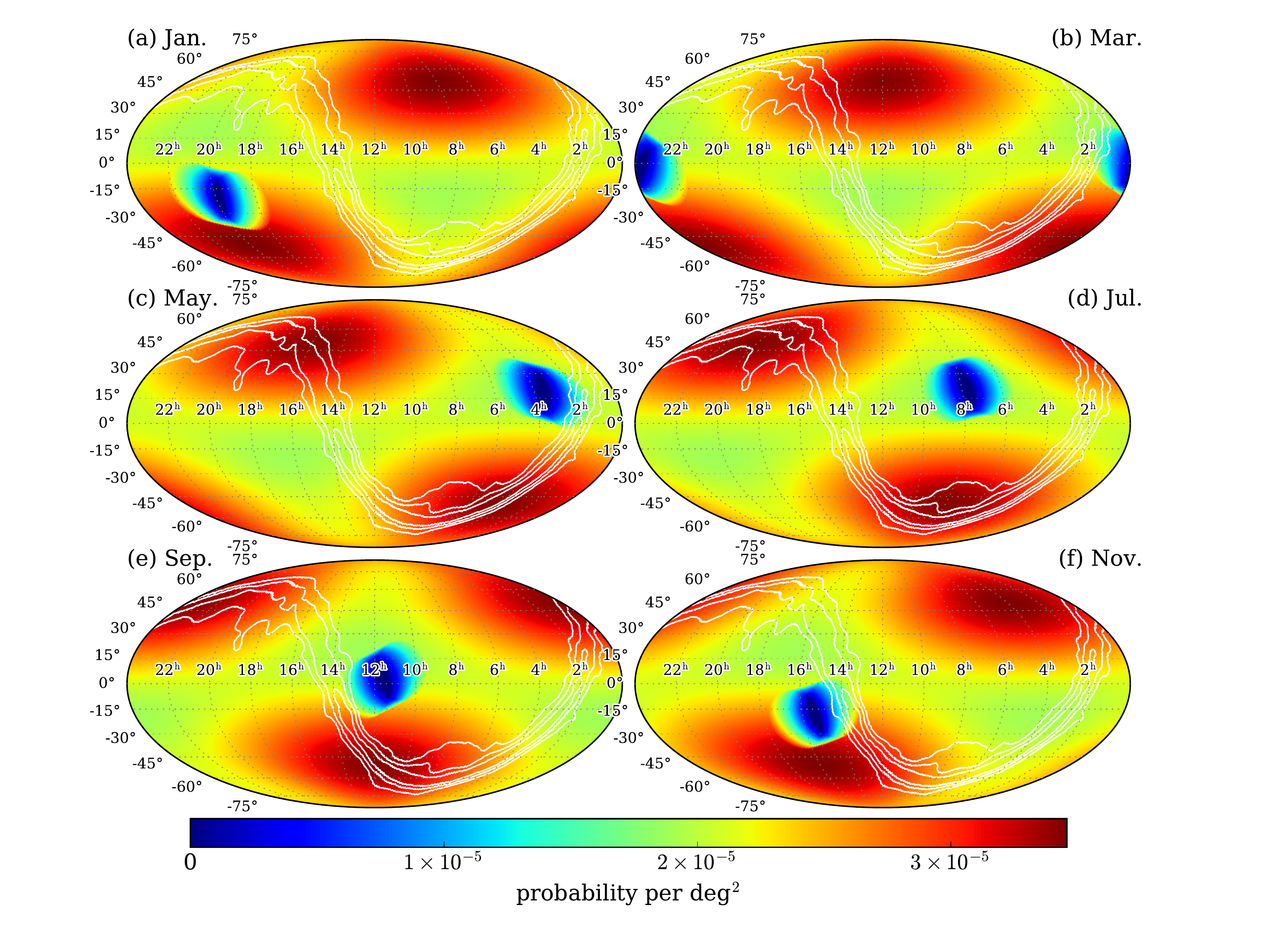}
    \caption{
        The monthly preferred regions for GW detections on the celestial sphere. 
        The Sun occludes a region 18$^\circ$ in radius (assuming observatories can see down to their horizons), and this region is smeared out by averaging over a month. 
        The white contours are the Planck occlusion masks~\citep{Planck} and represent the Galactic plane.
        }
    \label{f:seasonal bias}
\end{figure*}

The diurnal cycle preserves the dependence on right ascension as the Earth rotates (Fig \ref{f:gwbias}c), 
which persists over timescales of days to weeks. 
Figure~\ref{f:seasonal bias} shows the dependence averaged over a month for several months throughout the year. 
However, over the course of a year, the Earth's orbit will average away this dependence.
Nonetheless, as currently scheduled, GW detectors do not operate year-round, instead only recording data over a few consecutive months during observing runs.
If observing runs are scheduled during the same season repeatedly, the right ascension dependence introduced by the diurnal cycle can persist for years.

Furthermore, the galactic plane intersects the preferred directions from May through September (Fig.~\ref{f:seasonal bias}).
Depending on the desired target, this could be an advantage or a hindrance.
Detected compact binaries are expected to be extragalactic~\citep{detectionPRL,2016PhRvL.116x1103A} 
and the galactic plane will serve as a foreground for any EM follow-up, significantly complicating the removal of transient contaminants. 
However, detectable core-collapse Supernova are expected to be primarily galactic~\citep{Supernova}, and therefore GW detectors may have their best chance of observing such events during the North's summer.
Regardless of the source, a full understanding of the distribution of detectable signals across both declination and right ascension will be crucial when considering any isotropy or homogeneity measurements using GW observations alone.

\section{Impact on ground-based electromagnetic follow-up}\label{s:em}

Selection effects associated with ground-based GW detectors impact EM follow-up facilities.
In particular, we focus on three possible effects: the localization probability that a telescope can survey (observable probability), the source's closest angle of approach to an observatory's zenith while the observatory can observe (mimimum zenith distance), and the time until a GW source becomes observable (delay time).
We focus on EM follow-up timescales of up to a few days or weeks, and therefore neglect seasonal modulations of the sky over the duration of the EM observations for each individual GW event.
This is an appropriate timescale for short gamma-ray burst optical afterglows~\citep{sGRB} and kilonovae~\citep{kilonova}, two promising EM counterparts of compact binary coalescences involving at least one neutron star. 
Radio transients may persist over longer timescales and therefore these effects may be less relevant. 
For these EM follow-up timescales, we consider low-latency GW alerts. 
In O1, it took a couple of days to issue alerts, but we expect this to be reduced to a few minutes in O2 and beyond~\citep{emfollow}.
Throughout this paper, we assume the diurnal cycle modeled in Equation~\ref{e:bias} as well as 18$^\circ$ of astronomical twilight~\citep{2006A&A...455..385P}. 
We also assume the observatories can observe within 90$^\circ$ of their zenith.

In the limit of a large number of detections, the combined posterior distributions trace out the network antenna pattern. 
In what follows, we use the antenna pattern to approximate limiting distribution of many events.\footnote{Please see the Appendix for more details.}
To quantify departures from this limit when only a few events are available, we simulate collections of events based on binary black hole localizations with two detectors~\citep{BF2Y}. 
We find good agreement.


\subsection{Observable Probability}\label{s:obsprob}
GW localizations are driven by triangulation and, for networks of two detectors, the localization is characterized by large rings that can span hundreds of square degrees regardless of source morphology.
Furthermore, these error regions typically have support at antipodal points on the sky, making it difficult for a single EM observatory to access the entire skymap.
This is compounded by solar occlusion, which renders certain parts of the sky inaccessible.\footnote{We ignore lunar occlusion because, unlike solar occlusion, it is not thought to systematically correlate with when the detectors operate.}
Because we focus on timescales of a few days, we assume the Earth will revolve at least once.
Therefore, the observable region depends only on the observatory's latitude. 

\begin{figure*}[!b]
  \begin{minipage}{0.5\textwidth}
    \includegraphics[width=1.0\columnwidth]{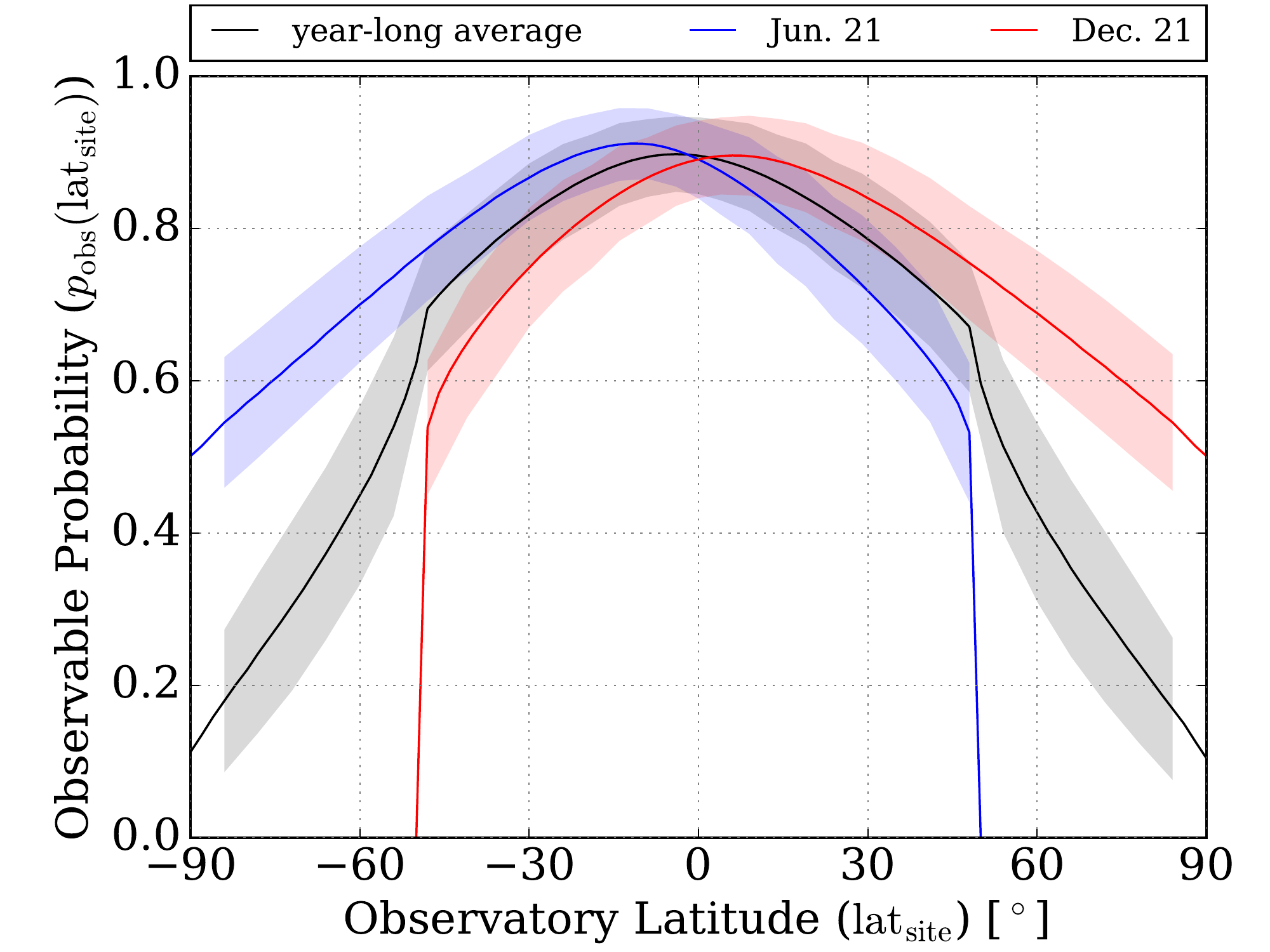}
    \caption{
        $p_\mathrm{obs}(\mathrm{lat_{site}})$ averaged over a year (black), near the Northern summer solstice (blue), and near the Northern winter solstice (red).
        Shaded regions correspond to the fluctuations from localization maps simulations ($\lim_{N_d\rightarrow\infty}(N_d/10)^{1/2}\sigma_{\hat{p}_\mathrm{obs}}$).
        }
    \label{f:obsprob}
  \end{minipage}
  \begin{minipage}{0.5\textwidth}
    \includegraphics[width=1.0\columnwidth]{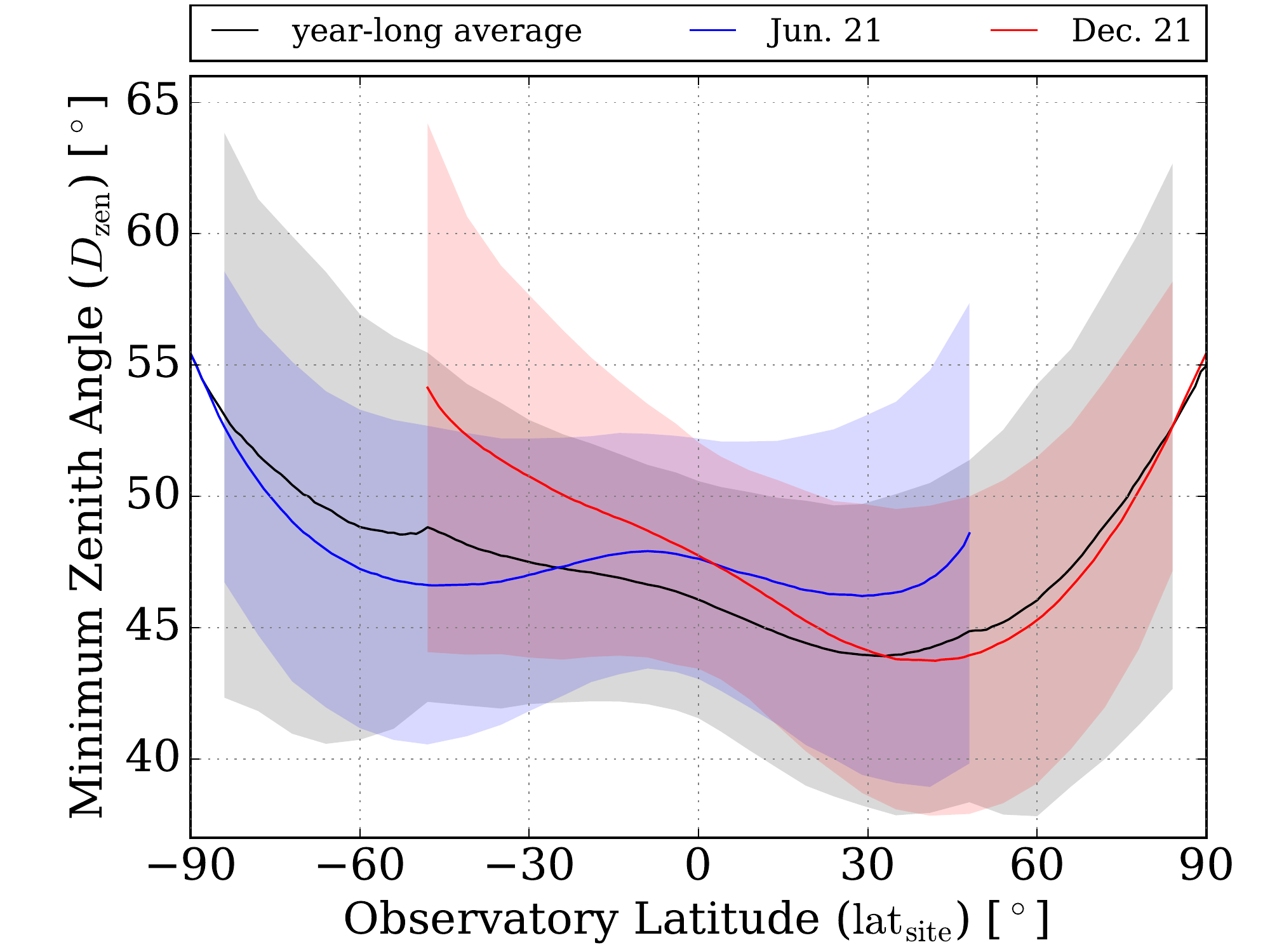}
    \caption{
        $D_\mathrm{zen}$ averaged over a year (black), near the Northern summer solstice (blue), and near the Northern winter soltice (red).
        Shaded regions correspond to the fluctuations from localization maps simulations ($\lim_{N_d\rightarrow\infty}(N_d/10)^{1/2}\sigma_{\hat{D}_\mathrm{zen}}$), and we note that $\hat{D}_\mathrm{zen}$'s distribution is quite broad.
        }
    \label{f:zenith}
  \end{minipage}
\end{figure*}

Figure~\ref{f:obsprob} shows the observable probability as a function of latitude for a year-long average and for the solstices.
We define
\begin{eqnarray}\label{e:pobs}
    p_\mathrm{obs}(\mathrm{lat_{site}}) &=& \int dt\, \bigg ( p(t|{\rm operating}) \nonumber \\
     & &\int d\Omega\, p_{GW}(\Omega,t) \Theta_{\mathrm{obs}}(\Omega,t,\mathrm{lat_{site}}) \bigg )
\end{eqnarray}
where $p_{GW}(\Omega,t)$ is the probability density of the GW source coming from $\Omega$ (at time $t$), and $\Theta_{\mathrm{obs}}$ is the region accessible from a particular latitude at $t$, respectively.
We see sharp declines near latitudes of $\pm50^\circ$ corresponding to the Arctic/Antarctic circles and astronomical twilight, but otherwise $p_\mathrm{obs}$ follows a smooth distribution favoring equatorial observatories. 
This is because equatorial observatories can systematically access sources in both hemispheres, whereas other observatories may be confined to only the probability within their own.
While equatorial facilities are favored overall, mid-latitudes have larger $p_\mathrm{obs}$ than they would for isotropically distributed detections.
This will persist regardless of the day/night cycle and is driven solely by the GW detectors' latitudes. We note that the optimal observing months of the Northern and Southern hemispheres are out of phase (Fig.~\ref{f:obsprob}). 
For example, if LIGO operates from September to February, the Northern hemisphere will have a better chance of observing counterparts than the Southern hemisphere.

While $p_\mathrm{obs}$ does not depend on the longitude, the diurnal cycle can still introduce a systematic bias.
This is because the Sun will be systematically out of phase with the Northern maximum of the antenna pattern (Fig.~\ref{f:seasonal bias}). 
We find that this typically produces an increase in $p_\mathrm{obs}$ of a few percent for mid-latitudes in the North, even though their analogs in the South have larger $p_\mathrm{obs}$ because of the shorter solar exposure over an entire year.

We note that $p_\mathrm{obs}$ reflects the amount of localization probability that is observable in the limit of many detections.
Outside of this limit, we consider an analogous quantity ($\hat{p}_\mathrm{obs}$) defined for a finite number of detections ($N_d$) with the correspondence $p_\mathrm{obs} = \lim_{N_d\rightarrow\infty} \hat{p}_\mathrm{obs}$.
\footnote{Please refer to Sec S1 and Sec S2.1 in the Appendix.}
Statistical fluctuations in $\hat{p}_\mathrm{obs}$, 
caused by variations in which events occur, can be large, particularly with $\lesssim$10 events.
We expect the uncertainty in this estimate to scale inversely with $\sqrt{N_d}$, 
and Figure~\ref{f:obsprob} reports $\lim_{N_d\rightarrow\infty} (N_d/10)^{1/2} \sigma_{\hat{p}_\mathrm{obs}}$ as error bars along with the mean.
Typical values are between 6--10\%. We also note that the intrinsic distribution for single events ($N_d=1$) 
may not be Gaussian but, in the limit of many detections, the distribution of the mean will be.


\subsection{Minimum Zenith Distance}\label{s:zenith}

While equatorial observatories may be able to access the largest integrated probability, and thereby have the largest probability of being able to image a source, this does not necessarily imply that they will have the best conditions for observing.
An important consideration is the closest approach of each field to an observatory's zenith (minimum zenith distance).
Sources at large angles from an observatory's zenith can be difficult to observe because of high airmass and mechanical limitations.
In Fig.~\ref{f:zenith} we present the mean minimum zenith distance, as a function of latitude, weighted by the probability that the source actually comes from each location:
\begin{eqnarray}\label{e:zenith}
    & &D_\mathrm{zen}(\mathrm{lat_{site}}) = \frac{1}{p_\mathrm{obs}}  \int dt\, \bigg ( p(t|{\rm operating}) \nonumber \\
     & & \int d\Omega \, p_{GW}(\Omega,t) \Theta_{\mathrm{obs}}(\Omega,t,\mathrm{lat_{site}}) D_\mathrm{zen}(\Omega, t, \mathrm{lat_{site}}) \bigg )
\end{eqnarray}
$\Theta_{\mathrm{obs}}$ accounts for solar occlusion and $D_\mathrm{zen}(\Omega, t, \mathrm{lat_{site}})$ incorporates when the EM facility will actually be able to observe.
We find that observatories at extreme latitudes ($\pm90^\circ$) will almost always have large $D_\mathrm{zen}$, with a gradual transition to lower values at lower latitudes. 
There is a $\sim10\%$ difference in $D_\mathrm{zen}$ between observatories at mid-latitudes, with Northern sites preferred.
This is because the diurnal cycle makes the Sun preferentially overlap the Southern antenna pattern and forces Southern facilities to observe closer to sunrise and sunset than their Northern counterparts. 
This behavior is particularly evident at the Northern winter solstice.

Fig.~\ref{f:zenith} also shows the fluctuations in the analgous stastic defined for a finite number of detections ($\hat{D}_\mathrm{zen}$).
We typically find $\lim_{N_d\rightarrow\infty}(N_d/10)^{1/2}\sigma_{\hat{D}_\mathrm{zen}}\sim 4$--10$^\circ$.\footnote{Please refer to Sec S1 and Sec S2.2 in the Appendix.}
Furthemore, $D_\mathrm{zen}$ corresponds to the mean of many events.
For a single event, the mode of $\hat{D}_\mathrm{zen}(N_d=1)$'s distribution falls near 20$^\circ$ at mid-latitudes in the North and 60$^\circ$ in the South, whereas the mode is near 50$^\circ$ at both poles.



\subsection{Delay Time}\label{s:delay}

For counterparts with timescales of days, EM observatories' longitudes can play an important role.
This is because observatories will have to wait to begin observing until the source rises at their location.
This can be exacerbated by the position of the Sun, which will systematically correlate with the diurnal cycle.

\begin{figure}
    \includegraphics[width=1.0\columnwidth]{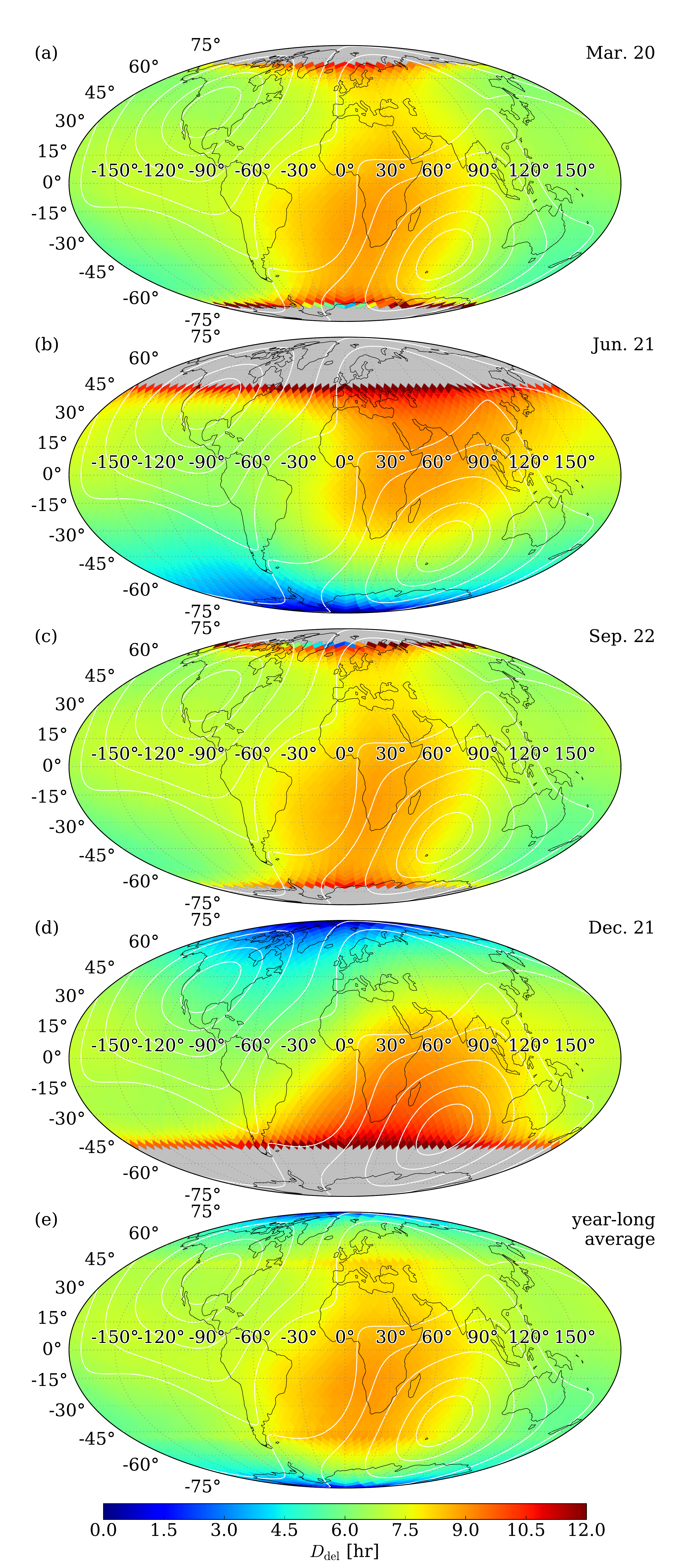}
    \caption{
        $D_\mathrm{del}(\Omega_\mathrm{site})$ for hypothetical observatories placed throughout the globe for times surrounding the Northern (a) spring equinox, (b) summer solstice, (c) fall equinox, and (d) winter solstice.
        (e) Shows a year-long average. 
        Grey regions correspond to observatories that are in perpetual daylight or twilight at the corresponding time.
        }
    \label{f:delay}
\end{figure}

We expect that the amount of time an observatory must wait before commencing observations (delay time) will depend on both the observatory's longitude and latitude and define
\begin{eqnarray}\label{e:delay}
    & &D_\mathrm{del}(\Omega_\mathrm{site}) = \frac{1}{p_\mathrm{obs}}  \int dt\, \bigg ( p(t|{\rm operating}) \nonumber \\ 
    & &\int d\Omega \, p_{GW}(\Omega,t) \Theta_{\mathrm{obs}}(\Omega,t,\mathrm{lat_{site}}) D_\mathrm{del}(\Omega, t, \Omega_\mathrm{site}) \bigg )
\end{eqnarray}
where $D_\mathrm{del}(\Omega_\mathrm{site}, t, \Omega)$ accounts for sunrise and sunset along with the source's relative position to the observatory.
We restrict ourselves to only the parts of the skymap that are actually accessible from each site.
Fig.~\ref{f:delay} shows the dependence on the observatory's longitude and latitude. 
Although the shape changes, we see a reasonably uniform distribution of $D_\mathrm{del}$ throughout the globe\footnote{There are slightly longer delay times in the North compared to their analogs in the South away from the blob over Africa.} with the notable exception of the Southern Atlantic, Africa, and the Indian Ocean, particularly in the year-long average (Fig \ref{f:delay}e).
Because the majority of detections will occur during North America's night, these locations are likely to already be in daylight and will therefore have to wait for sunset.
Furthermore, observatories in the South will have to wait until their zenith is very close to the source's right ascension before they observe, which explains why the blob is wider in the South.
From simulations outside the limit of many detections, we also note that the distribution of the analogous statistic ($\hat{D}_\mathrm{del}(N_d=1)$) is very non-Gaussian and very skew right. 
However, the mean is still Gaussian as $N_d\rightarrow\infty$ with an associated variance 
$\lim_{N_d\rightarrow\infty}(N_d/10)^{1/2}\sigma_{\hat{D}_\mathrm{del}} \sim$ 80--240 minutes, with most values near 100 minutes, 
depending on the observatory's location.\footnote{Please refer to Sec S1 and Sec S2.3 in the Appendix.}

We note that if the diurnal cycle were to be strongly reduced or eliminated, the blob above Africa would become less prominent and an analogous blob would appear in the Northern Pacific.
The latter corresponds to observatories that are in daylight when a source is detected above the Southern antenna pattern.
The overall effect of removing the diurnal cycle, however, is toward a more uniform distribution of delay times across the entire globe.

\section{Discussion}\label{s:discussion}

In this letter, we enumerate three effects that imprint selection effects on the distribution of GW detections, which will be relevant for all types of GW sources: (i) the detectors locations introduce a preference for mid-declinations, (ii) a diurnal cycle modulates when the detectors operate and produces a sky sensitivity with a dependence on the right ascension over short time-scales, and, (iii) if detections are made only during relatively short observing runs, the right ascension bias can be imprinted over longer timescales. 
While the effects are important when modeling the expected distribution of GW detections on the sky, they can also have significant implications for EM follow-up.

In general, ground-based EM observatories located at latitudes comparable to those of the GW detectors are preferred. 
While equatorial sites can access the largest fraction of the sky, mid-latitude sites in the North will be able to observe more events closer to their zenith without sacrificing much coverage.
Furthermore, there is a preference for EM observatories located near the same longitude as the GW detectors, or slightly West thereof, assuming the EM facilities can begin observing quickly and that counterparts decay with timescales of hours.
This leads us to the conclusion that EM facilities located near the GW detectors are favored by how the GW detectors' actually operate.

In addition to informing follow-up efforts for GW candidates, our results may help guide EM activities \textit{in advance} of GW observations. 
For example, because we know which parts of the celestial sphere are most likely to host detectable GW events, we can build up relevant templates for image subtraction before GW observations even begin, focusing particularly on the highest probability regions. 
Galaxy catalog constructions could be focused similarly. 
In addition, surveys may focus their observations near the peak of the instantaneous antenna pattern, thereby anticipating where detectable GW sources are most likely to occur and facilitating target of opportunity follow-ups.
This will also increase the probability of serendipitous detection of prompt EM counterparts.
We note that EM observatories located near GW detectors will naturally survey a maximum of the antenna pattern because their zenith lies near that maximum automatically. 

The behavior of GW observatories can become more complicated as the global network of detectors expands. 
A larger network increases the uniformity of sensitivity to GW signals across the sky, and therefore reduces the selection effects in both declination and right ascension.
Furthermore, detectors located around the globe will likely experience diurnal cycles that are out of phase, further reducing any preference for certain right ascensions.
However, because all planned detectors lie within a relatively confined band of latitudes~\citep{0264-9381-32-2-024001,PhysRevD.88.043007,1742-6596-484-1-012007}, the bias on declination may persist at some level.
In addition, the LIGO detectors will likely have the best sensitivity in the network for at least the next few years~\citep{ObservingScenarios}.
This means that they will generally provide the dominant contribution to detections, and the observed distribution of sources will follow their antenna pattern. 
Thus, our analysis serves as a reasonable prediction for both the distributions of GW detections as well as their impact on EM follow-up efforts for the next few years.


\section{Acknowledgements}

The authors want to thank Lisa Barsotti, Matt Evans, and Rich Middleman for discussions about locking behavior, commissioning, 
and anthropogenic noise sources, James Annis for discussions about EM observations, and Marica Branchesi for very helpful comments. 
They also acknowledge the support of the National Science Foundation and the LIGO Laboratory. 
LIGO was constructed by the California Institute of Technology and Massachusetts Institute of Technology with funding from the National Science Foundation and operates under cooperative agreement PHY-0757058. 
The authors would like to acknowledge the LIGO Data Grid clusters. Specifically, we thank the Albert Einstein Institute in Hannover, supported by the Max-PlanckGesellschaft, for use of the Atlas high-performance computing cluster.

DEH and HYC were supported by NSF CAREER grant PHY-1151836. They also acknowledge support
from the Kavli Institute for Cosmological Physics at the University of Chicago through 
NSF grant PHY-1125897 as well as an endowment from the Kavli Foundation.

This is LIGO document P1600219.

\newpage

\appendix
\renewcommand\thefigure{S\arabic{figure}}    
\renewcommand\theequation{S\arabic{equation}}    
\renewcommand\thesection{S\arabic{section}}    
 
We provide the details of the calculations and the distributions of the finite number statistics in the paper. 
We also derive our model of the diurnal cycle. 
\enspace


\section{calculating means in the limit of many detections}\label{s:direct calculation}

Throughout the paper, we use the antenna pattern to approximate limiting distributions of many events. 
We also study the distributions by drawing simulated localization maps. 
We repeatedly draw $N_d$ maps, distributed through time according to $p(t|\mathrm{operating})$, to obtain the distributions of our statistics for $N_d$ detections. In this approach 
the observable probability, the minimum zenith distance, and the delay time are calculated as:
\begin{equation}
    \hat{p}_\mathrm{obs}(\mathrm{lat_{site}}) = \frac{1}{N_d} \sum\limits_{i=1}^{N_d} \int d\Omega\, p_{{\rm sky},i} \Theta_{\mathrm{obs}}(\Omega,t_i,\mathrm{lat_{site}}) \label{eqn:pobs_map} 
\end{equation}
\begin{equation}
    \hat{D}_\mathrm{zen}(\mathrm{lat_{site}}) = \frac{1}{N_d\, \hat{p}_\mathrm{obs}} \sum\limits_{i=1}^{N_d} \int d\Omega\, p_{{\rm sky},i} \Theta_{\mathrm{obs},i} D_\mathrm{zen}(\Omega, t_i, \mathrm{lat_{site}}) \label{eqn:zenith_map} 
\end{equation}
\begin{equation}
    \hat{D}_\mathrm{del}(\Omega_\mathrm{site}) = \frac{1}{N_d\, \hat{p}_\mathrm{obs}} \sum\limits_{i=1}^{N_d} \int d\Omega\, p_{{\rm sky},i} \Theta_{\mathrm{obs},i} D_\mathrm{del}(\Omega, t_i, \Omega_\mathrm{site}) \label{eqn:delaytime_map} 
\end{equation}
where $p_{{\rm sky},i}$ is the localization map probability for the $i_{\rm th}$ detection and $\hat{p}_\mathrm{obs}$ in Equations~\ref{eqn:zenith_map} and~\ref{eqn:delaytime_map} is computed using the same set of skymaps as the explicit sum.
We note that using the antenna pattern yields the mean of these statistics when simulating many detections.
To wit, if we calculate $\chi$ using the antenna pattern and $\hat{\chi}$ using sets of simulated maps, we expect
\begin{eqnarray}
  \chi & = & \lim\limits_{N_d\rightarrow\infty} \hat{\chi} \nonumber \\
       & = & \lim\limits_{N_d\rightarrow\infty} \frac{1}{N_d}\sum\limits_{i=1}^{N_d} \int d\Omega\, p_{{\rm sky},i} \chi(t_i) 
\end{eqnarray}
We note that $\chi(t_i)$ only depends on the time the event occurs through the position of the Sun; the dependence will be the same for all detections that occur at the same time.
If we break the sum into small segments of time, we can write
\begin{eqnarray}
  \chi & = & \lim\limits_{N_d\rightarrow\infty} \frac{1}{N_d}\sum\limits_{j=1}^{N_t} \sum\limits_{\substack{i=1 \\ t_i\in[t_j,t_j+\Delta t)}}^{N_{d,j}} \int d\Omega\, p_{{\rm sky},i} \chi(t_i) \\
  \nonumber
\end{eqnarray}
where $\sum_{j} N_{d,j} = N_d$. Now, when $N_d\rightarrow\infty$, we can make the segments as small as we like while maintaining a large number of detections in each bin.
We then obtain
\begin{widetext}
\begin{eqnarray}
    \chi & = & \lim\limits_{\substack{N_t\rightarrow\infty\\N_t\Delta t = T}} \sum\limits_{j=1}^{N_t} \lim\limits_{N_{d,j} \rightarrow\infty} \frac{1}{N_d} \sum\limits_{\substack{i=1 \\ t_i\in[t_j,t_j+\Delta t)}}^{N_{d,j}} \int d\Omega\, p_\mathrm{sky,i} \chi(t_i) \nonumber \\
         & = & \lim\limits_{\substack{N_t\rightarrow\infty\\N_t\Delta t = T}} \sum\limits_{j=1}^{N_t} \lim\limits_{N_{d,j} \rightarrow\infty} \left( \frac{N_{d,j}}{N_d} \right) \int d\Omega \left( \frac{1}{N_{d,j}} \sum\limits_{\substack{i=1 \\ t_i\in[t_j,t_j+\Delta t)}}^{N_{d,j}} p_\mathrm{sky,i} \right) \chi(t_i) \nonumber \\
         & = & \lim\limits_{\substack{N_t\rightarrow\infty\\N_t\Delta t = T}} \sum\limits_{j=1}^{N_t} \left( p(t|\mathrm{operating})\Delta t \right) \int d\Omega \left( p_{GW}(\Omega, t_j) \right) \chi(t_j) \nonumber \\
         & = & \int dt\, p(t|\mathrm{operating}) \int d\Omega\, p_{GW}(\Omega, t) \chi(t) 
\end{eqnarray}
\end{widetext}
where $T$ is the length of the observing season. We assumed $t_i\rightarrow t_j$ for all events within each bin and used the fact that many localization posteriors stacked on top of one another will average to the network antenna pattern in Equatorial coordinates ($p_{GW}(t_j)$), which depends on the bin's time.
We have also used the fact that the fraction of events occuring in each bin is equal to the probability that the detectors are operating throughout that bin ($p(t|\mathrm{operating})\Delta t$).
By approximating these integrals, we obtain the limits of the means much more efficiently than through direct simulation.
This proceedure, or an equivalent, is used in Figures~\ref{f:obsprob},~\ref{f:zenith}, and~\ref{f:delay}.

We also note that we normalize by the total observable probability in Equations~\ref{e:zenith}, \ref{e:delay}, \ref{eqn:zenith_map}, and \ref{eqn:delaytime_map}.
This is because we restrict ourselves to only the fraction of the probability that is actually observable.
The particular form of our normalization (dividing by $\hat{p}_\mathrm{obs}$ for a set of events rather than each event separately) guarantees that we sample the antenna pattern in the limit.


\section{distributions of statistics for individual events}\label{s:indiv distrib}

Section~\ref{s:em} presents the limit of our statistics when many detections are present. 
In this limit, the distributions of these statistics will be Gaussian, but it is also informative to examine the distributions when there are only a few events.
This is particularly important when considering how an observatory could be impacted for any individual event, rather than for a collection of events.
We also consider these distributions in the limit of extremely well localized sources: $p_{{\rm sky},i}\rightarrow\delta(\Omega-\Omega_i)$.
This point source limit describes the distributions obtained when only observing the location of the true source.

These distributions were computed using bootstrapped simulations of BBH detections based off~\citet{BF2Y}.
The library developed to perform these simulations is publicly available (\url{https://github.com/reedessick/selectionEffects}) and readers are encouraged to use it to determine distributions for their favorite observatory.


\subsection{$\hat{p}_\mathrm{obs}$}

\begin{figure*}
  \begin{minipage}{0.5\textwidth}
    \includegraphics[width=1.0\columnwidth]{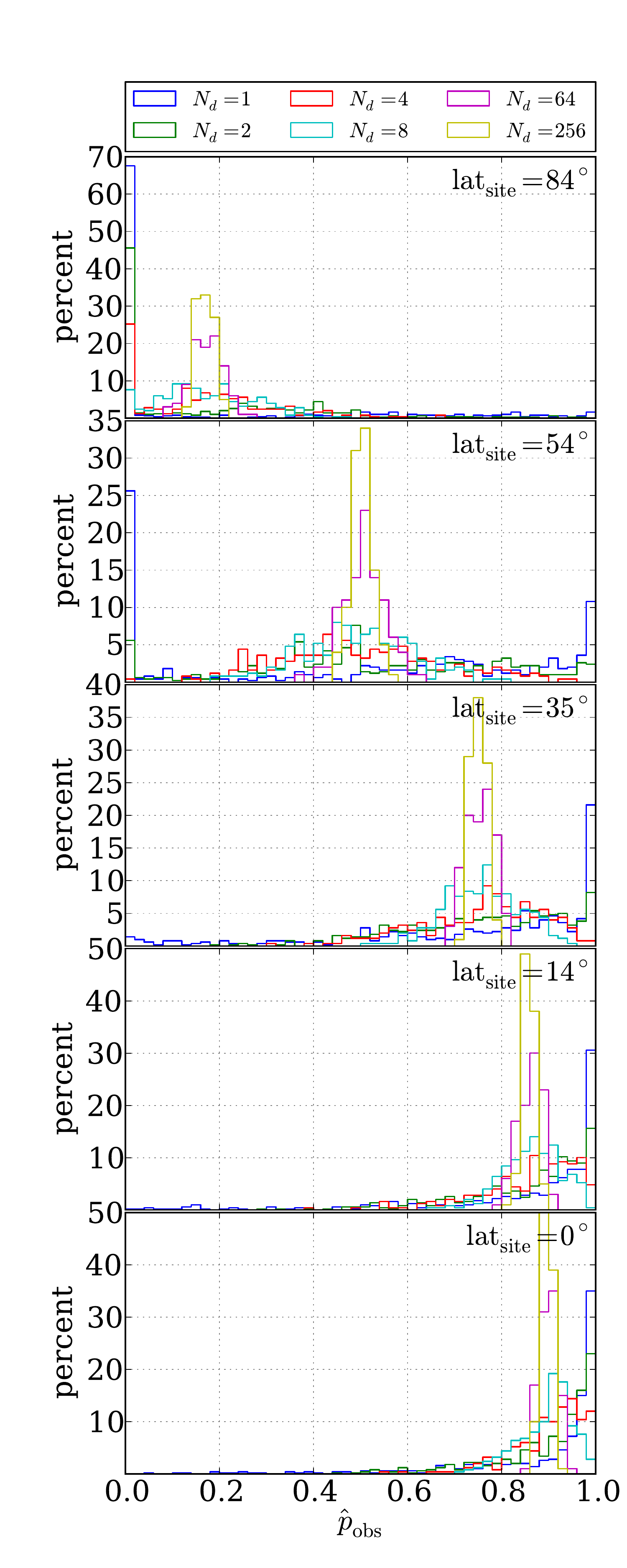}
    \caption{
        Distributions of $\hat{p}_\mathrm{obs}$ for a few latitudes and $N_d$ for year-long averages.
        We note that the distributions with $N_d=1$ may not be very Gaussian and display long tails.
        Nonetheless, as $N_d\rightarrow\infty$, the means of the distributions tend toward the values reported in Figure~\ref{f:obsprob}.
        }
    \label{f:pobs distrib}
  \end{minipage}
  \begin{minipage}{0.5\textwidth}
    \includegraphics[width=1.0\columnwidth]{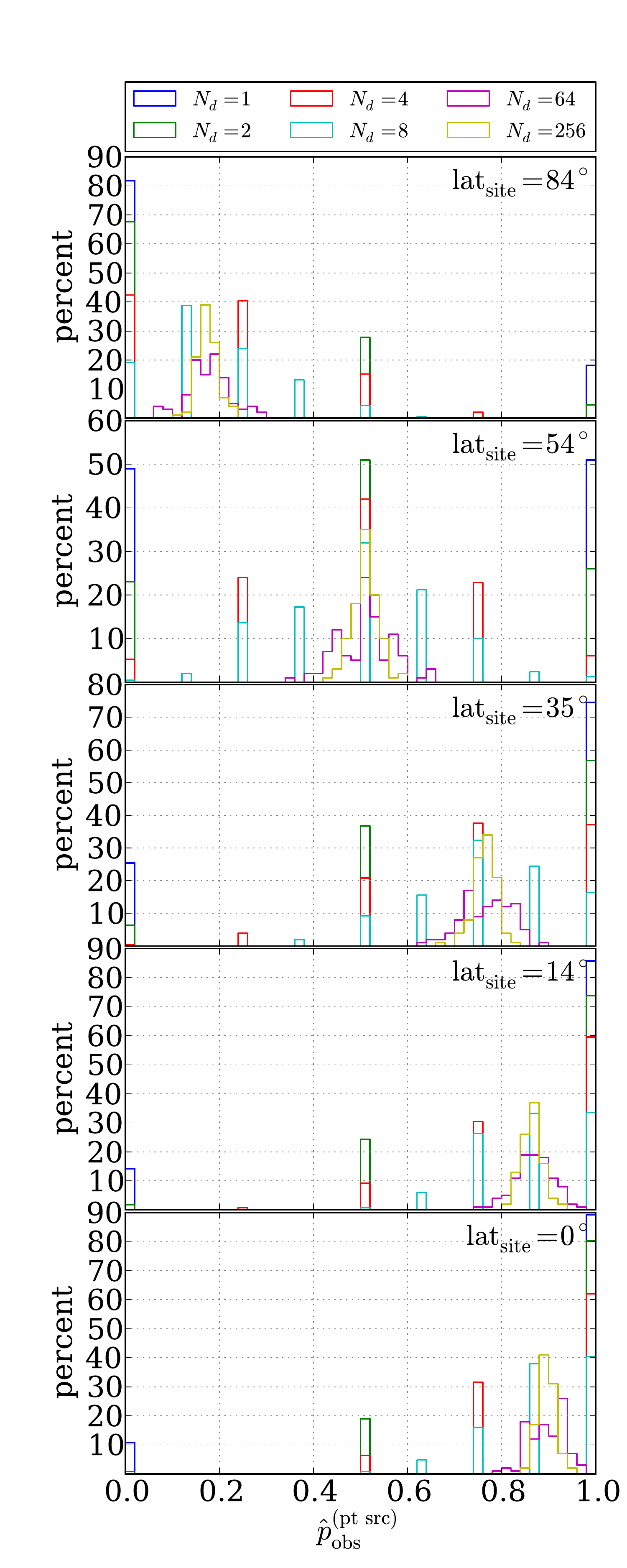}
    \caption{
        Distributions of the fraction of observable point sources ($\hat{p}^{(\mathrm{pt\,src})}_\mathrm{obs}$) for a few latitudes and $N_d$.
        We note that the distributions tend toward means equivalent to those in Figure~\ref{f:pobs distrib} as $N_d\rightarrow\infty$, although their discreteness for small $N_d$ somewhat obscures this.
        }
    \label{f:pobs point distrib}
  \end{minipage}
\end{figure*}

Of all the statistics we consider, $\hat{p}_\mathrm{obs}$ is the most Gaussian for small $N_d$ for year-long averages.
However, because it is bounded from above and below, the distribution does deviate at times.
Figure~\ref{f:pobs distrib} shows the distributions for a few latitudes and a few values of $N_d$. 

We note that, in the limit $N_d\rightarrow\infty$, $p_\mathrm{obs}$ is the fraction of true counterparts that an observatory can observe. 
When considering extremely well localized sources, the observatory will either be able to observe the true source or not, and each trial will have nearly the same probability of success, modulo variations caused by the Sun's declination.
Therefore, the fraction of true sources each observatory can observe will be nearly binomially distributed.
Figure~\ref{f:pobs point distrib} plots the point-source limit of these distributions ($\hat{p}^{(\mathrm{pt\,src})}_\mathrm{obs}$), 
again with several values of $N_d$.


\subsection{$\hat{D}_\mathrm{zen}$}

\begin{figure*}
  \begin{minipage}{0.5\textwidth}
    \includegraphics[width=1.0\columnwidth]{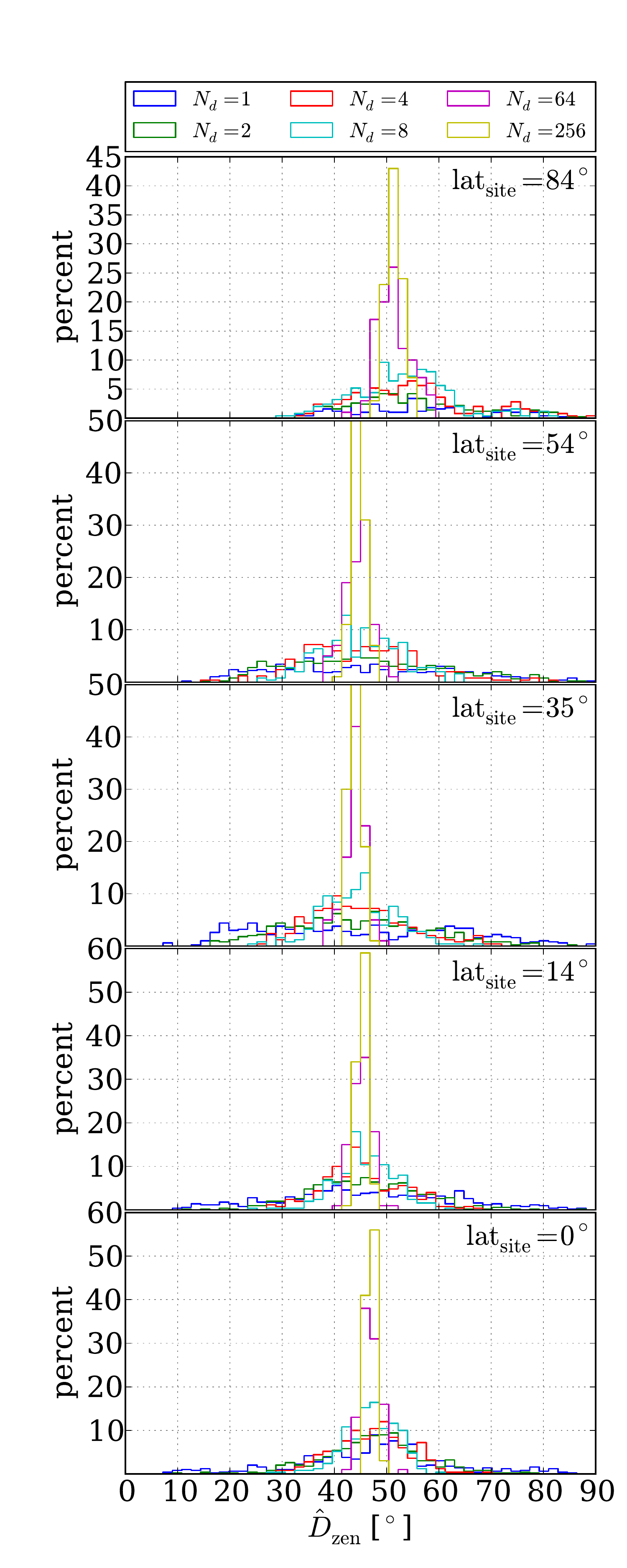}
    \caption{
        Distributions of $\hat{D}_\mathrm{zen}$ for a few latitudes and $N_d$ for year-long averages. 
        We note that the distributions are rather broad and all centered near similar values.
        This is reflected in the wide error bars and similar means in Figure~\ref{f:zenith}.
        Nonetheless, the mean of the distribution collapses to the same values as in Figure~\ref{f:Dzenith point distrib} as $N_d\rightarrow\infty$.
        }
    \label{f:Dzenith distrib}
  \end{minipage}
  \begin{minipage}{0.5\textwidth}
    \includegraphics[width=1.0\columnwidth]{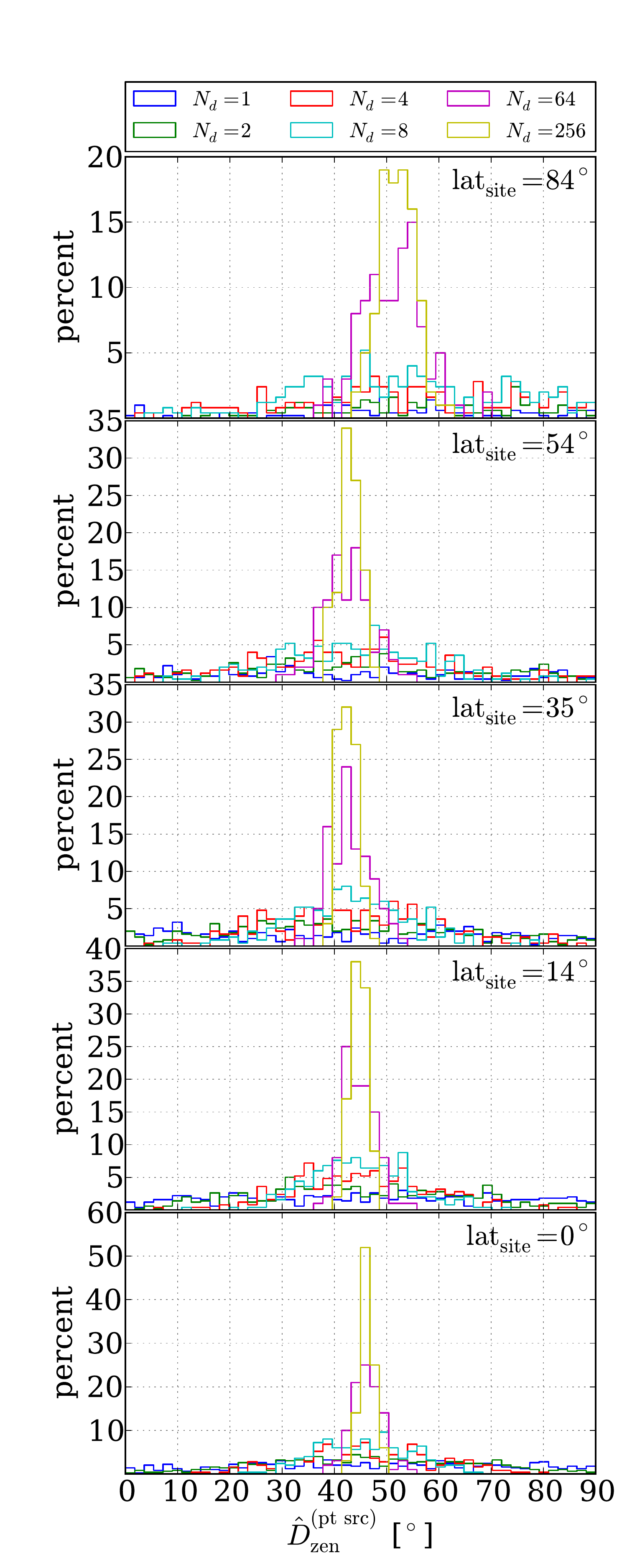}
    \caption{
        Distributions of $\hat{D}_\mathrm{zen}^{(\mathrm{pt\,src})}$ for a few latitudes and $N_d$ for year-long averages.
        We note that these distributions have more shot noise than those in Figure~\ref{f:Dzenith distrib}, evident in the modes of the mid-latitude distributions.
        $\hat{D}_\mathrm{zen}(N_d=1)$ is often larger than $\hat{D}^{(\mathrm{pt\,src})}_\mathrm{zen}(N_d=1)$ because the triangulation rings can reach accross most of the antenna pattern with only two detectors.
        }
    \label{f:Dzenith point distrib}
  \end{minipage}
\end{figure*}

$\hat{D}_\mathrm{zen}(N_d=1)$ is also fairly Gaussian for some latitudes, but there can often be non-trivial deviations therefrom. 
In particular, mid-latitudes may show interesting skew right distributional shapes.
Figure~\ref{f:Dzenith distrib} shows these distributions for a few latitudes as a function of $N_d$.
We note that these distributions are much narrower than the point source limit in which all events are well localized ($\hat{D}^{(\mathrm{pt\,src})}_\mathrm{zen}$), shown in Figure~\ref{f:Dzenith point distrib}.
There is more ``shot noise'' for $\hat{D}^{(\mathrm{pt\,src})}_\mathrm{zen}$, which broadens the distributions, but the general distributional shapes are similar for both $\hat{D}_\mathrm{zen}$ and $\hat{D}^{(\mathrm{pt\,src})}_\mathrm{zen}$.


\subsection{$\hat{D}_\mathrm{del}$}

\begin{figure*}
  \begin{minipage}{0.5\textwidth}
    \includegraphics[width=1.0\columnwidth]{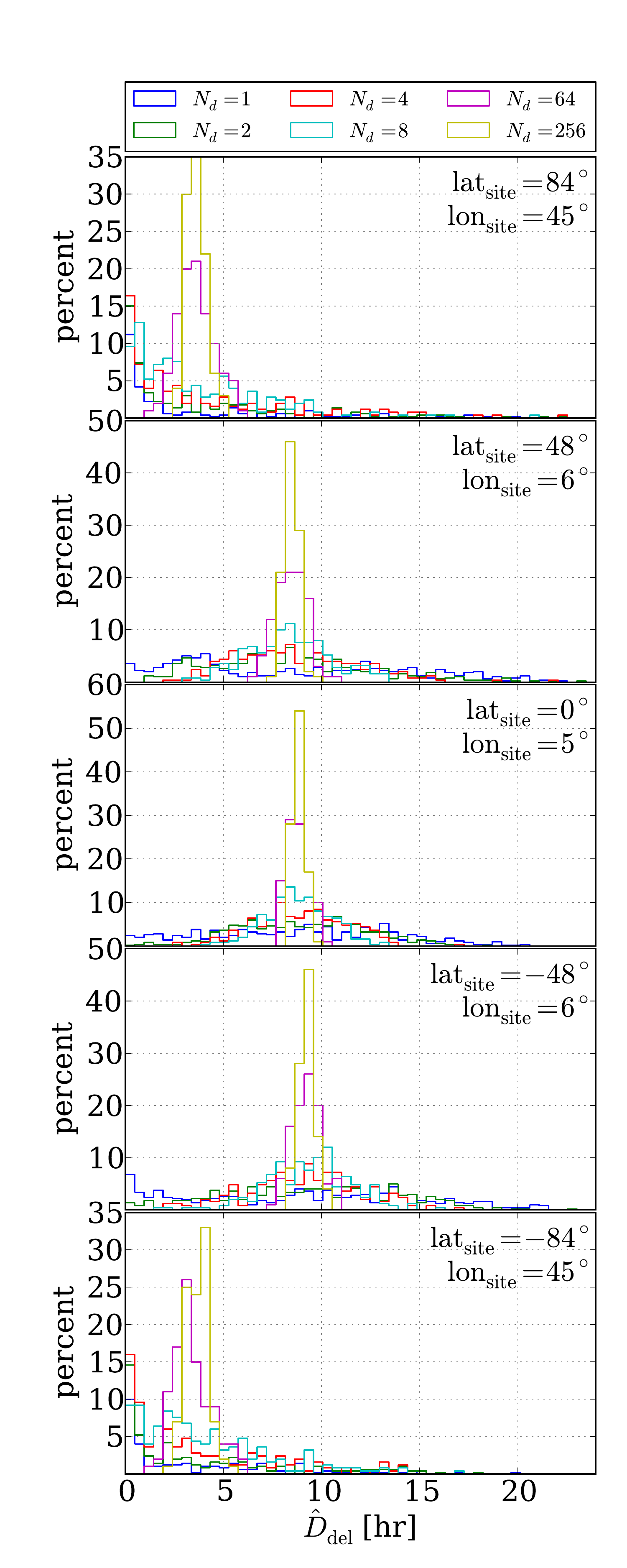}
    \caption{
        Distributions of $\hat{D}_\mathrm{del}$ for a few sites and $N_d$ for year-long averages.
        We note that the $N_d=1$ distributions have large modes near $\hat{D}_\mathrm{del}=0$, corresponding to events that are immediately observable, as well as very broad support extending to long $\hat{D}_\mathrm{del}$.
        This is particular evident near the poles.
        }
    \label{f:Ddelay distrib}
  \end{minipage}
  \begin{minipage}{0.5\textwidth}
    \includegraphics[width=1.0\columnwidth]{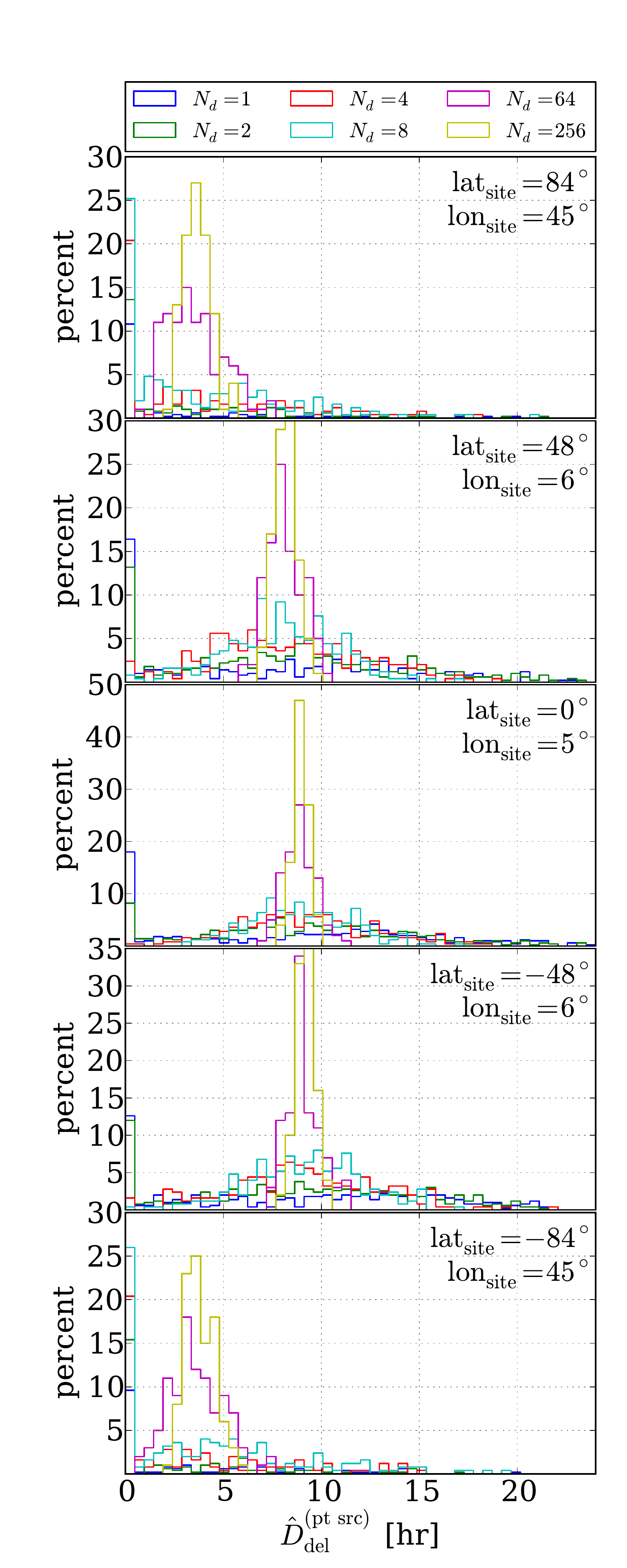}
    \caption{
        Distributions of $\hat{D}_\mathrm{del}^{(\mathrm{pt\,src})}$ for a few sites and $N_d$ for year-long averages.
        The modes near $\hat{D}_\mathrm{del}^{(\mathrm{pt\,src})}(N_d=1)$ are more pronounced than in Figure~\ref{f:Ddelay distrib} because there is more shot noise in the point source measurement.
        In fact, averaging over the skymap tends to broaden the distributions for $N_d\lesssim5$.
        }
    \label{f:Ddelay point distrib}
  \end{minipage}
\end{figure*}

We note in \S~\ref{s:delay} that the distribution of $\hat{D}_\mathrm{del}$ is very skew right when $N_d$ is small.
Figure~\ref{f:Ddelay distrib} demonstrates this.
Typically, there is an extremely large lobe near zero, corresponding to events that are immediately observable, and a long tail comprised of events that require waiting.
When we consider the point source limit of well localized events, this behavior is enhanced.
Figure~\ref{f:Ddelay point distrib} demonstrates that.
When we integrate over typical skymaps, instead of point sources, we find that the peak is smeared out to longer delay times.
This is because different parts of the skymap may become observable at different times, and that fuzz tends to smooth the distribution.

\section{derivation of model for $p(t|\mathrm{operating})$}\label{s:derivation of p(t|u)}

While Equation~\ref{e:bias} is fairly self evident from inspection of Figure~\ref{f:gwbias}, we can derive its form from more basic assumptions about how human activity may cause downtime.
We posit two states of a network of detectors: \textit{up} ($u$) and \textit{down} ($d$), with science-quality data available only in the \textit{up} state.
Furthermore, we posit two possible causes for detectors being in the \textit{down} state: \textit{random} causes ($r$) that are uncorrelated with time and \textit{cyclic} causes ($c$) which are correlated with time, usually through a diurnal cycle. 
We note that the \textit{random} and \textit{cyclic} models are not necessarily mutually exclusive and the detector could be down for mutliple reasons at the same time.

Therefore, we have
$$
  p(d|t) = p(d|c,t)p(c|t) + p(d|r,t)p(r|t) - p(d|c \cap r, t)p(c \cap r|t)
$$
which implies 
\begin{widetext}
$$
  p(d) = \int\, dt p(t) p(d|t) = \int dt\, p(t) \left[ p(d|c,t)p(c|t) + p(d|r,t)p(r|t) - p(d|c \cap r, t)p(c \cap r|t) \right]
$$
\end{widetext}
Furthermore, because the \textit{up} and \textit{down} states are mutually exclusive and span the space of possible detector states at any single time, we have
$$
  p(u|t) = 1 - p(d|t) \Rightarrow p(u) = 1 - p(d)
$$
and Bayes theorem yields
$$
  p(t|u) = \frac{p(u|t)p(t)}{p(u)} = \frac{\left( 1 - p(d|t) \right)p(t)}{1 - p(d)}
$$

We measure $p(u)$ and $p(t|u)$ in a straightforward manner from the data used in Figure~\ref{f:gwbias}.
Typically, we assume some periodicity in $p(t|u)$ to generate a histogram with enough samples to be statistically meaningful, but we expect to be able to identify the periodic elements of $p(t|u)$ through Fourier analysis as well.

However, we are also interested in slightly different probabilities. To wit, we would like to know
$$
  p(c|d,t) = \frac{p(d|c,t)p(c|t)}{p(d|t)}
$$
$$
  p(r|d,t) = \frac{p(d|r,t)p(r|t)}{p(d|t)}
$$
and
$$
  p(c \cap r |d,t) = \frac{p(d|c \cap r, t)p(c \cap r|t)}{p(d|t)}
$$
which express the probabilities that the detector is down due to a particular cause at a specific time. 
If we only care about long-term averages (over time-scales much longer than the \textit{cyclic} model's periodicty), then we can marginalize away the time dependence:
$$
  p(c|d) = \int\, dt p(t) \frac{p(d|c,t)p(c|t)}{p(d|t)}
$$
$$
  p(r|d) = \int\, dt p(t) \frac{p(d|r,t)p(r|t)}{p(d|t)}
$$
and
$$
  p(c \cap r |d) = \int\, dt p(t) \frac{p(d|c \cap r, t)p(c \cap r|t)}{p(d|t)}
$$
These equations should hold regardless of the specific form of the \textit{cyclic} and \textit{random} models.

By definition, we assume that the \textit{random} causes are uncorrelated with time so that $p(d|r,t) = p(d|r)$. 
Furthermore, we assume some periodic function for the \textit{cyclic} model so that $p(d|c,t) = p(d|c,t+\tau)$ for some $\tau$. 
Specifically, we expand the periodic function in terms of the oscillating (AC) and constant (DC) components
$$
  p(d|c,t) = p_{DC}(d|c) + p_{AC}(d|c,t)
$$
such that

$$
  p_{DC} = \int\limits_{0}^{\tau} dt\, p(d|c,t)
$$
Clearly, we require $p_{DC}(d|c) \geq -p_{AC}(d|c,t)\, \forall\, t$. 
We also assume the priors for the causes do not depend on time ($p(r|t)=p(r)$ and $p(c|t)=p(c)$) and that the priors are equal for the two causes ($p(r)=p(c)=p$).

Note: by measuring $p(d)$, we only extract the combination of $p_{DC}(d|c) + p(d|r)$ and cannot separate these terms further.
However, by measuring $p(t|u)$ as well, we are able to determine $p_{AC}(d|c,t)$ from which we can determine $p_{DC}(d|c)$ by requiring that $\min_{t}\left\{p(d|c,t)\right\} = 0$.
Any other DC component to the \textit{cyclic} model is indistinguishable from the \textit{random} model and therefore we lump it together with the \textit{random} model.

If we allow the cause models to overlap but require them to be independent, we obtain $p(d|c \cap r, t) = p(d|c,t) p(d|r,t)$.
This implies 
$$
  p(r) + p(c) - p(r)p(c) = 1 \Rightarrow p(r)=p(c)=p=1
$$
and the two observables are 
$$
  p(d) = p(d|r)p + \left(1-p(d|r)p\right) p_{DC}(d|c)p
$$
\begin{widetext}
\begin{eqnarray*}
  p(t|u) & = & p(t) \frac{1 - p(d|r)p - (1-p(d|r)p)p_{DC}(d|c)p - (1-p(d|r)p)p_{AC}(d|c,t)p}{1 - p(d|r)p - (1-p(d|r)p)p_{DC}(d|c)p} \\
         & = & p(t)\left( 1 - \frac{(1-p(d|r)p)p}{1 - p(d)} p_{AC}(d|c,t) \right)
\end{eqnarray*}
\end{widetext}
This yields
\begin{widetext}
\begin{eqnarray*}
  p(c|d,t) = \frac{p_{AC}(d|c,t) + p_{DC}(d|c)}{\left(p_{AC}(d|c,t) + p_{DC}(d|c)\right)\left(1 - p(d|r)p\right) + p(d|r)}
\end{eqnarray*}
\begin{eqnarray*}
  p(r|d,t) = \frac{p(d|r)}{\left(p_{AC}(d|c,t) + p_{DC}(d|c)\right)\left(1 - p(d|r)p\right) + p(d|r)}
\end{eqnarray*}
\begin{eqnarray*}
  p(c \cap r|d,t) = \frac{\left(p_{AC}(d|c,t) + p_{DC}(d|c)\right)p(d|r)p}{\left(p_{AC}(d|c,t) + p_{DC}(d|c)\right)\left(1 - p(d|r)p\right) + p(d|r)}
\end{eqnarray*}
\end{widetext}
\begin{widetext}
and their marginalized counterparts
\begin{eqnarray*}
  p(c|d) & = & \int dt\, p(t|d) \frac{p_{AC}(d|c,t) + p_{DC}(d|c)}{\left(p_{AC}(d|c,t) + p_{DC}(d|c)\right)\left(1 - p(d|r)p\right) + p(d|r)} 
          =  \frac{p_{DC}(d|c)p}{p(d)}
\end{eqnarray*}
\begin{eqnarray*}
  p(r|d) & = & \int dt\, p(t|d) \frac{p(d|r)}{\left(p_{AC}(d|c,t) + p_{DC}(d|c)\right)\left(1 - p(d|r)p\right) + p(d|r)} 
          =  \frac{p(d|r)p}{p(d)}
\end{eqnarray*}
\begin{eqnarray*}
  p(c \cap r|d) & = & \int dt\, p(t|d) \frac{\left(p_{AC}(d|c,t) + p_{DC}(d|c)\right)p(d|r)p}{\left(p_{AC}(d|c,t) + p_{DC}(d|c)\right)\left(1 - p(d|r)p\right) + p(d|r)} 
                 =  \frac{p_{DC}(d|c)p \cdot p(d|r)p}{p(d)}
\end{eqnarray*}
\end{widetext}
A reasonable ansatz is $p(d|c,t) = B\left(1+\sin\left(\frac{2\pi t}{\tau} - \phi\right)\right)$, in which case we obtain
$$
  p(d) = p(d|r)p + Bp(1-p(d|r)p)
$$
\begin{eqnarray}
  p(t|u) & = & p(t)\left( 1 + \frac{(1-p(d|r)p)pB}{1-p(d)}\sin\psi\right) \nonumber \\
         & = & p(t)\left( 1 + A\sin\psi\right) \nonumber
\end{eqnarray}
which is exactly the form of Equation~\ref{e:bias}.
We also see why a low duty cycle (small $p(u)=1-p(d)$) can amplify the amplitude of they day/night bias.

If we are interested in just the probability associated with the \textit{cyclic} model regardless of the \textit{random} model, we obtain
$$
  p(c|d) = \frac{1-p(d)}{p(d)}\left(\frac{A}{1-p(d)+(1-p(d))A}\right) \approx 0.49
$$
which implies that nearly half the time \textit{cyclic} causes were at least partly responsible for bringing down the detector network during O1.
If we restrict ourselves to times when the \textit{cyclic} causes were the sole cause of the downtime, we obtain 
$$
  p(c|d) - p(c\cap r|d) = \frac{(1-p(d))A}{p(d)} \approx 0.29
$$
which suggests that we could reduce the downtime by 30\% if we completely removed \textit{cyclic} causes of downtime.


\bibliography{refs}

\end{document}